\documentclass[prd,twocolumn,showpacs,amsmath,amssymb]{revtex4-1}

\usepackage{mathrsfs}
\usepackage{amssymb}
\usepackage{amsmath}
\usepackage{epsfig}
\usepackage{graphicx}

\usepackage{subfigure}
\usepackage{hyperref}
\usepackage{multirow}
\usepackage{overpic}
\usepackage{color}
\usepackage{ulem}

\usepackage{lineno}
\usepackage{dcolumn}
\usepackage{bm}

\newcommand{\jpsi}{J/\psi}

\newcommand{\pppipi}{p \overline p \pi^+ \pi^-}

\newcommand{\DsRhom}{D_s^-\rho^+}

\newcommand{\antiDKst}{\overline{D}{}^0\overline{K}{}^{\ast0}}

\begin{document}

\title{\boldmath Search for the rare decays $\jpsi \to \DsRhom$ and
$\jpsi \to \antiDKst$} \date{\today}

\author{\small
M.~Ablikim$^{1}$, M.~N.~Achasov$^{8,a}$, X.~C.~Ai$^{1}$, O.~Albayrak$^{4}$, M.~Albrecht$^{3}$, D.~J.~Ambrose$^{41}$, F.~F.~An$^{1}$, Q.~An$^{42}$, J.~Z.~Bai$^{1}$, R.~Baldini Ferroli$^{19A}$
% ...
\\
\vspace{0.2cm}
(BESIII Collaboration)\\
\vspace{0.2cm} {\it
$^{1}$ Institute of High Energy Physics, Beijing 100049, People's Republic of China\\
% ...
$^{3}$ Bochum Ruhr-University, D-44780 Bochum, Germany\\
$^{4}$ Carnegie Mellon University, Pittsburgh, Pennsylvania 15213, USA\\
% ...
$^{8}$ G.I. Budker Institute of Nuclear Physics SB RAS (BINP), Novosibirsk 630090, Russia\\
% ...
$^{19}$ (A)INFN Laboratori Nazionali di Frascati, I-00044, Frascati, Italy; (B)INFN and University of Perugia, I-06100, Perugia, Italy\\
% ...
$^{41}$ University of Rochester, Rochester, New York 14627, USA\\
$^{42}$ University of Science and Technology of China, Hefei 230026, People's Republic of China\\
% ...
$^{a}$ Also at the Novosibirsk State University, Novosibirsk, 630090, Russia\\
% ...
}}

\begin{abstract}

 A search for the rare decays of $\jpsi \rightarrow \DsRhom + c.c.$ and
$\jpsi \rightarrow \antiDKst + c.c.$ is performed with a data sample of 225.3
million $\jpsi$ events collected with the BESIII detector. No evident
signal is observed. Upper limits on the branching fractions are
determined to be $\mathcal{B}(\jpsi \rightarrow \DsRhom + c.c.) < 1.3 \times
10^{-5}$ and $\mathcal{B}(\jpsi \rightarrow \antiDKst + c.c.) < 2.5 \times
10^{-6}$ at the 90\% confidence level.

\end{abstract}


\maketitle

\section{Introduction}
The decays of the low-lying charmonium state $\jpsi$, which is below
open-charm threshold, are dominated by strong interactions through
intermediate gluons and electromagnetic interactions through virtual
photons, where both the intermediate gluons and photons are produced by $c\bar{c}$ annihilation. However, flavor-changing weak decays of $\jpsi$ through virtual intermediate bosons
are also possible in the standard model (SM) framework, and the
branching fractions of $\jpsi$ inclusive weak decays are estimated to be
on the order of $10^{-8}$~\cite{sm}. Several models addressing new physics,
including the top-color model, the minimal supersymmetric standard
model (MSSM) with {\it R}-parity violation and a general
two-Higgs-doublet model (2HDM), allow $\jpsi$ flavor-changing
processes to occur with branching fractions around $10^{-5}$, which may
be measurable in experiments~\cite{zhangx2,zhangx1}. Searches for rare
$\jpsi$ decays to a single charmed meson provide an experimental test
of the SM, and a way to look for possible new physics beyond the SM.

The BESII experiment has searched for semileptonic decays and hadronic
decays of $\jpsi \to D^-_s \pi^+$, $\jpsi \to D^- \pi^+$, and $\jpsi
\to \overline{D}{}^0 \overline{K}{}^0$~\cite{cc}, and set upper limits on
the order of $10^{-4}{\sim}10^{-5}$ using a sample of $5.8\times10^7$
$\jpsi$ events~\cite{sunss,zhangjy}. With the prospect of high
statistics $\jpsi$ samples, theoretical calculations of the branching
fractions of two-body hadronic weak decays of $\jpsi \to DP/DV$, where
$D$ represents a charmed meson and $P$ and $V$  the pseudoscalar and vector mesons, respectively, have been performed ~\cite{rc,dhir,lihb,kk,wangym,wangym2}. The branching fractions of $\jpsi
\rightarrow D_s^-\rho^+$ and $\jpsi \rightarrow \antiDKst$ are
predicted to be higher than those of $\jpsi \rightarrow D_s^-\pi^+$
and $\jpsi \rightarrow \overline{D}{}^0\overline{K}{}^0$, e.g. the relative ratio ${\mathcal{B}(\jpsi \rightarrow D_s^-\rho^+)}/{\mathcal{B}(\jpsi \rightarrow D_s^-\pi^+)}=4.2$~\cite{kk}.

\begin{figure}[hbtp]
  \centering
  \subfigure[]{
    \label{feynman_diag:a}
    \includegraphics[scale=0.4]{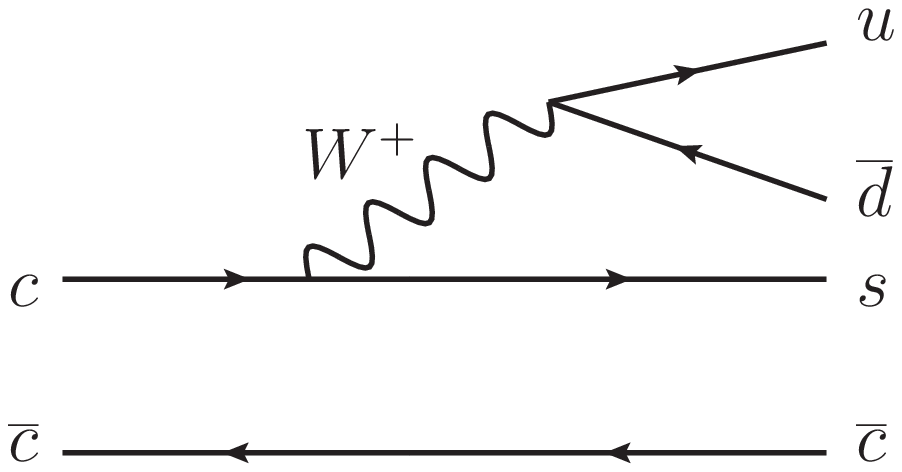}}~~~
  \subfigure[]{
    \label{feynman_diag:b}
    \includegraphics[scale=0.4]{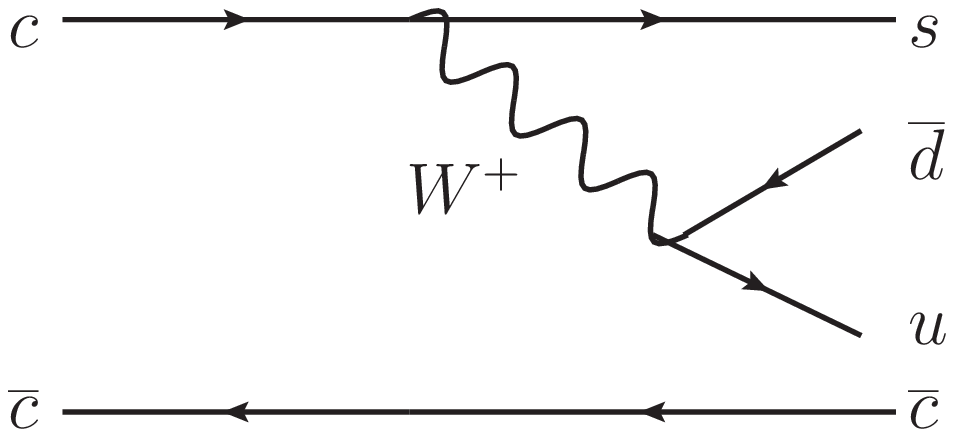}}~~
    \caption{Leading order Feynman diagrams for (a) $\jpsi \to D_s^-\rho^+$ and (b) $\jpsi \to \antiDKst$.}
  \label{feynman_diag}
\end{figure}

In this analysis, we search for two Cabibbo-favored decay modes
$\jpsi \rightarrow D_s^-\rho^+$ [Fig.~\ref{feynman_diag}(a)] and
$\jpsi \rightarrow \antiDKst$ [Fig.~\ref{feynman_diag}(b)] based on
($225.3 \pm2.8)\times 10^6$ $\jpsi$ events~\cite{jpsino} accumulated
with the Beijing Spectrometer III (BESIII) detector~\cite{bes3},
located at the Beijing Electron-Positron Collider
(BEPCII)~\cite{bepc}.

\section{The BESIII Experiment and Monte Carlo Simulation}

The BESIII detector with a geometrical acceptance of $93\%$ of $4\pi$, consists of: a small-celled, helium-based main drift chamber (MDC), an electromagnetic calorimeter (EMC) made of CsI(Tl) crystals, a plastic scintillator time-of-flight system (TOF), a super-conducting solenoid magnet, and a muon chamber system (MUC) made of Resistive Plate Chambers (RPCs). The detector has been described in detail elsewhere~\cite{bes3}.

The optimization of the event selection and the estimation of physics
backgrounds are performed using Monte Carlo (MC) simulated data
samples. The {\sc geant4}-based simulation software {\sc
boost}~\cite{boost} includes the geometric and material description of
the BESIII detectors and the detector response and digitization
models, as well as the tracking of the detector running conditions and
performance. The production of the $\jpsi$ resonance is simulated by
the MC event generator {\sc kkmc}~\cite{kkmc}; the known decay modes
are generated by {\sc evtgen}~\cite{evtgen} with branching fractions
set at world average values~\cite{pdg}, while the remaining unknown
decay modes are modeled by {\sc lundcharm}~\cite{lundcharm}.

\section{Data Analysis}
In order to avoid large background contamination from conventional
$\jpsi$ hadronic decays, the $D_s^-$ and $\overline{D}{}^0$
mesons are identified by their semileptonic decays $D_s^- \to \phi e^-
\overline{\nu}_e$ with $\phi \to K^+K^-$ and $\overline{D}{}^0 \to K^+
e^- \overline{\nu}_e$, where the electron is used to tag the events
and the missing energy due to the escaping neutrino is also used
to suppress backgrounds.  Since the neutrinos are undetectable, the $D_s^-$ and $\overline{D}{}^0$ mesons can not be directly identified by their invariant mass of the decay products. However, because of the two-body
final states, they can be identified in the distribution of mass
recoiling against the $\rho^+$ and $\overline{K}{}^{*0}$ in $\rho^+ \to \pi^+ \pi^0 (\pi^0 \to \gamma\gamma)$ and $\overline{K}{}^{*0} \to K^- \pi^+$ decays, respectively.

\begin{figure*}[hbtp]
  \centering
  \subfigure{
    \label{mass:a}
    \includegraphics[scale=0.4]{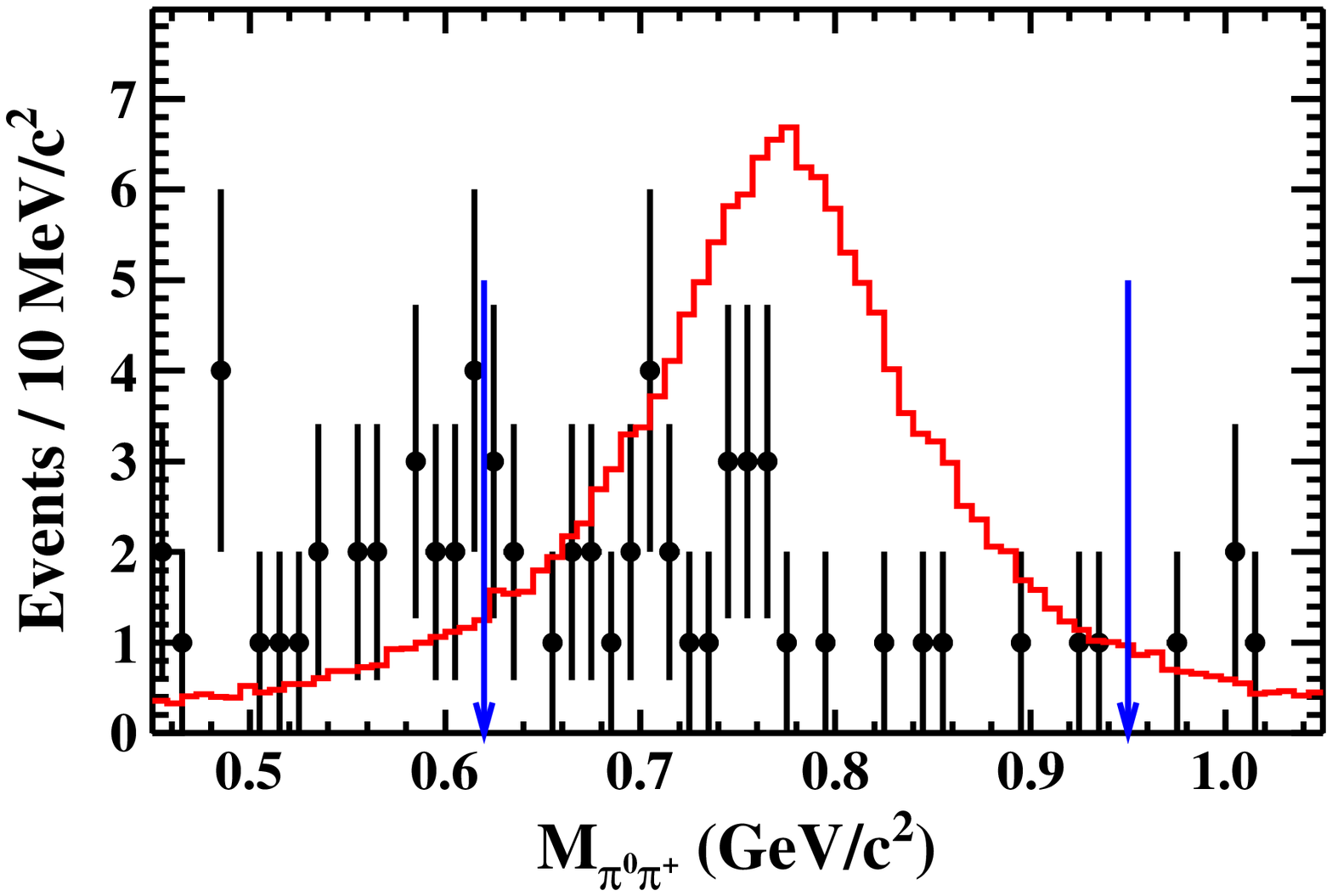}
    \put(-40,120){\bf (a)}}
  \subfigure{
    \label{mass:b}
    \includegraphics[scale=0.4]{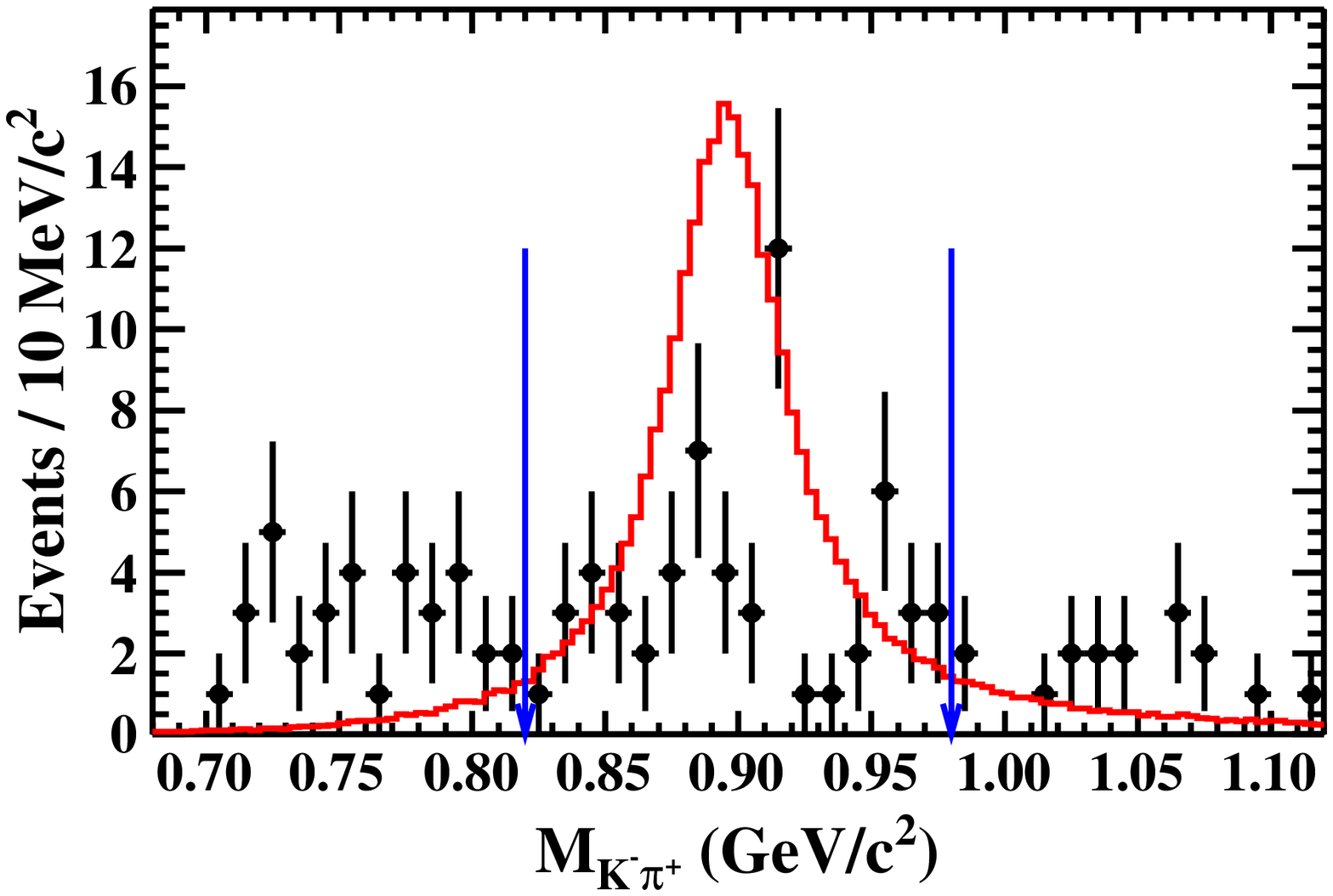}
    \put(-40,120){\bf (b)}}
    \caption{The invariant mass distributions of resonance candidates
    for (a) $\rho^+$ from $\jpsi \to \DsRhom$, $\rho^+ \to \pi^+
    \pi^0 (\pi^0 \to \gamma\gamma)$, and (b) $\overline{K}{}^{*0}$ from
    $\jpsi \to \antiDKst$, $\overline{K}{}^{*0} \to K^- \pi^+$.  The
    requirements of $M_{\pi^0\pi^+} \in (0.62,0.95)$\,GeV/$c^2$ and
    $M_{K^-\pi^+} \in (0.82,0.98)$\,GeV/$c^2$ are shown in the figures by vertical arrows. The dots with error bars are data, while the histograms represent distributions of the arbitrarily normalized exclusive signal MC events.}

  \label{mass}
\end{figure*}

\begin{figure*}[hbtp]
  \centering
  \subfigure{
    \label{nuep:a}
    \includegraphics[scale=0.4]{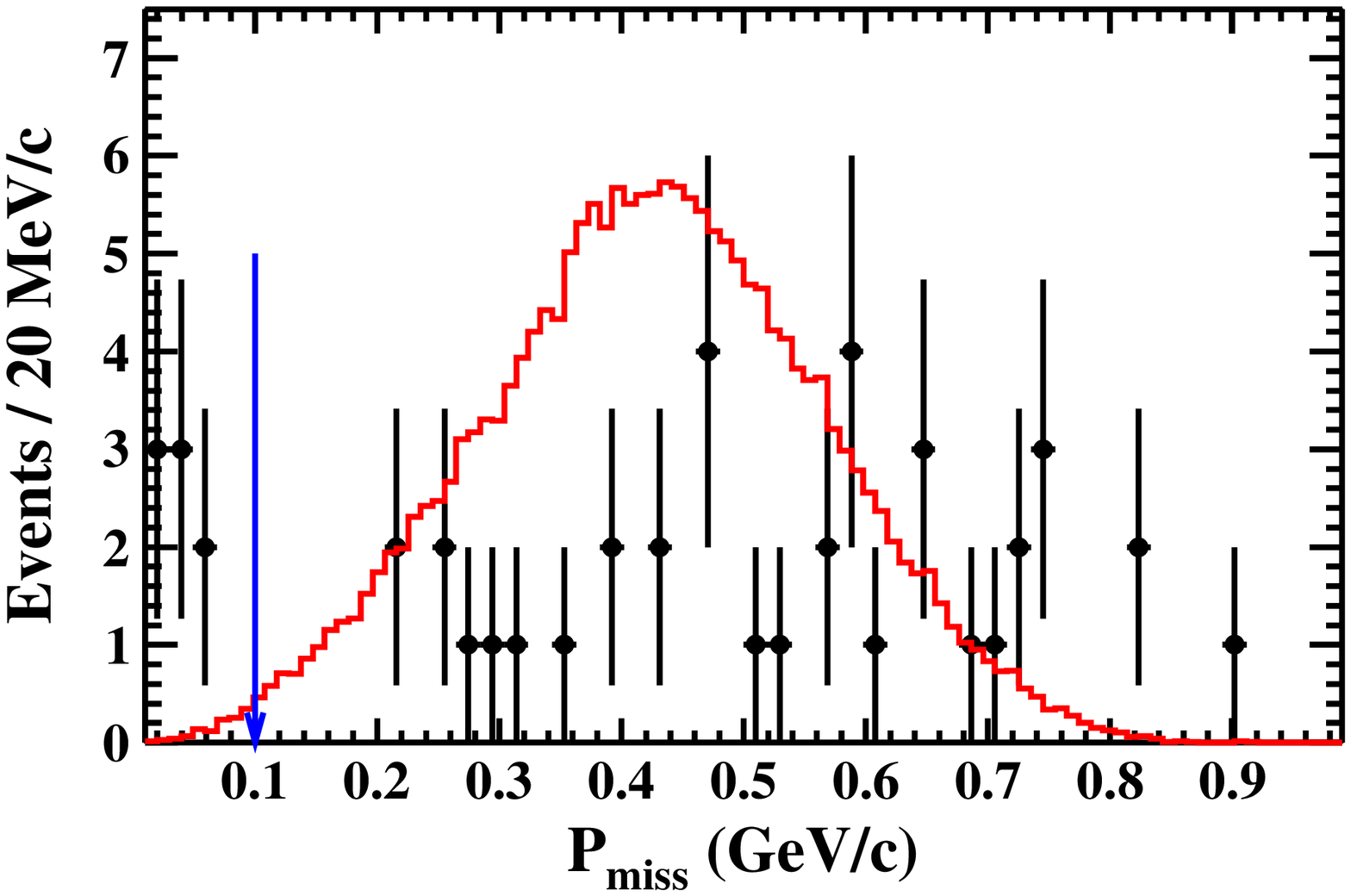}
    \put(-40,120){\bf (a)}}
  \subfigure{
    \label{nuep:b}
    \includegraphics[scale=0.4]{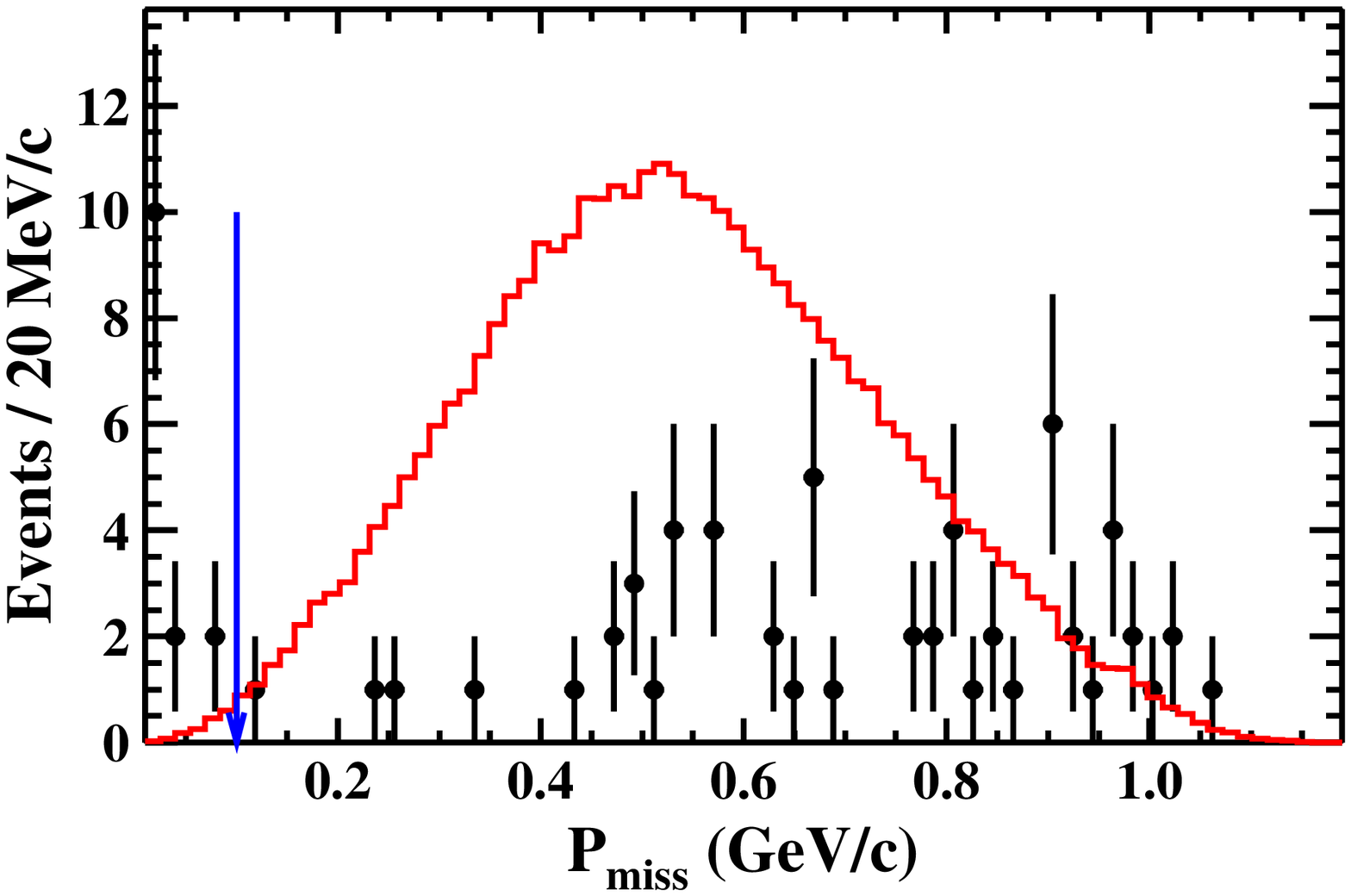}
    \put(-40,120){\bf (b)}}
    \caption{$P_{miss}$ distributions for the decay of (a)
    $\jpsi \to \DsRhom$, and (b) $\jpsi \to \antiDKst$.  The
    requirement $P_{miss}> 0.1$\,GeV/$c$ is shown in the
    figures by vertical arrows.  The dots with error bars are data, while the histograms represent distributions of the arbitrarily normalized exclusive signal MC events.}

  \label{nuep}
\end{figure*}

Charged tracks in BESIII are reconstructed from MDC hits. For each
charged track, the polar angle must satisfy $|\cos\theta |<
0.93$, and it must pass within $\pm 20$\,cm from the interaction point
in the beam direction and within $\pm 2$\,cm of the beam line in the
plane perpendicular to the beam. The number of charged tracks is
required to be four with zero net charge. The TOF and the specific
energy loss $dE/dx$ of a particle measured in the MDC are combined to
calculate particle identification (ID) probabilities Prob($i$), where $i (i
= e/\pi/K/p)$ is the particle type. Prob($K$) $>$ Prob($\pi$) and
Prob($K$) $>$ Prob($p$) are required for kaon candidates, while
Prob($\pi$) $>$ Prob($e$), Prob($\pi$) $>$ Prob($K$) and Prob($\pi$)
$>$ Prob($p$) are required for pion candidates. For electron
candidates, besides the particle identification requirement of
Prob($e$) $>$ Prob($\pi$) and Prob($e$) $>$ Prob($K$), $E/cP>0.8$ is
also required, where $E/cP$ is the ratio of the energy deposited in the
EMC to the momentum reconstructed from the MDC.  In addition,
$|\cos\theta|<0.8$ is required for electron candidates since the
particle ID efficiencies between data and MC agree better in the
barrel.

Photon candidates are reconstructed by clustering EMC crystal
energies. Efficiency and energy resolution are improved by including
energy deposits in nearby TOF counters. A photon candidate has to be
more than $20^{\circ}$ away from any charged track, and the minimum
energy is 25\,MeV for barrel showers $(|\cos\theta |< 0.80)$
and 50\,MeV for end cap showers $(0.86 < |\cos\theta |< 0.92)$.
An EMC timing requirement, i.e., $0 \le t \le 700$\,ns, is used to
suppress electronic noise and energy deposits in the EMC unrelated to
the events.  Kinematic fits of pairs of photon candidates to the
$\pi^0$ mass are performed. When there are more than two photons, all
possible $\gamma\gamma$ combinations are considered, and the one
yielding the smallest $\chi^2_{\gamma\gamma}$ is retained.

In the selection of $\jpsi \to \DsRhom \to \phi e^- \overline{\nu}_e
\pi^+ \pi^0 \to \gamma\gamma K^+K^-\pi^+e^-\overline{\nu}_e$, four
charged track candidates and at least two photons are required.
The invariant mass of $K^+K^-$ for a $\phi$ candidate is required to satisfy
$M_{K^+K^-} \in (1.01,1.03)$\,GeV/$c^2$. The invariant mass
distribution of $\rho^+ (\pi^0\pi^+)$ candidates is shown in
Fig.~\ref{mass:a}~\cite{plot}, and the requirement 0.62\,GeV/$c^2 <
M_{\pi^0\pi^+} < 0.95$\,GeV/$c^2$ is used to select $\rho$
candidates. The $\chi^2_{\gamma\gamma}$ of the kinematic fit should be less
than 200 for the $\pi^0$ candidates in this selection.

\begin{figure*}[hbtp]
  \centering
  \subfigure{
    \label{nueu:a}
    \includegraphics[scale=0.4]{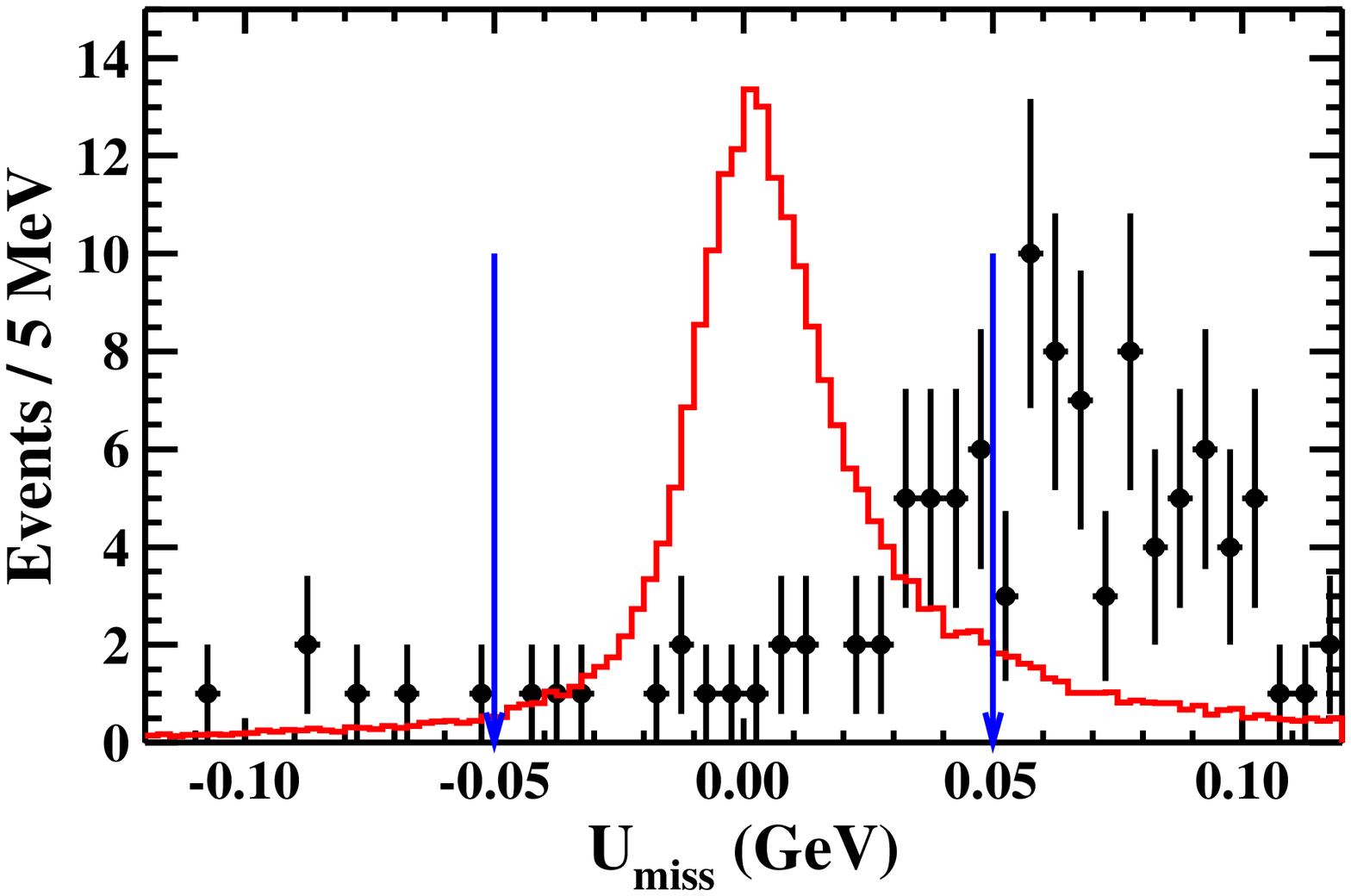}
    \put(-180,120){\bf (a)}}
  \subfigure{
    \label{nueu:b}
    \includegraphics[scale=0.4]{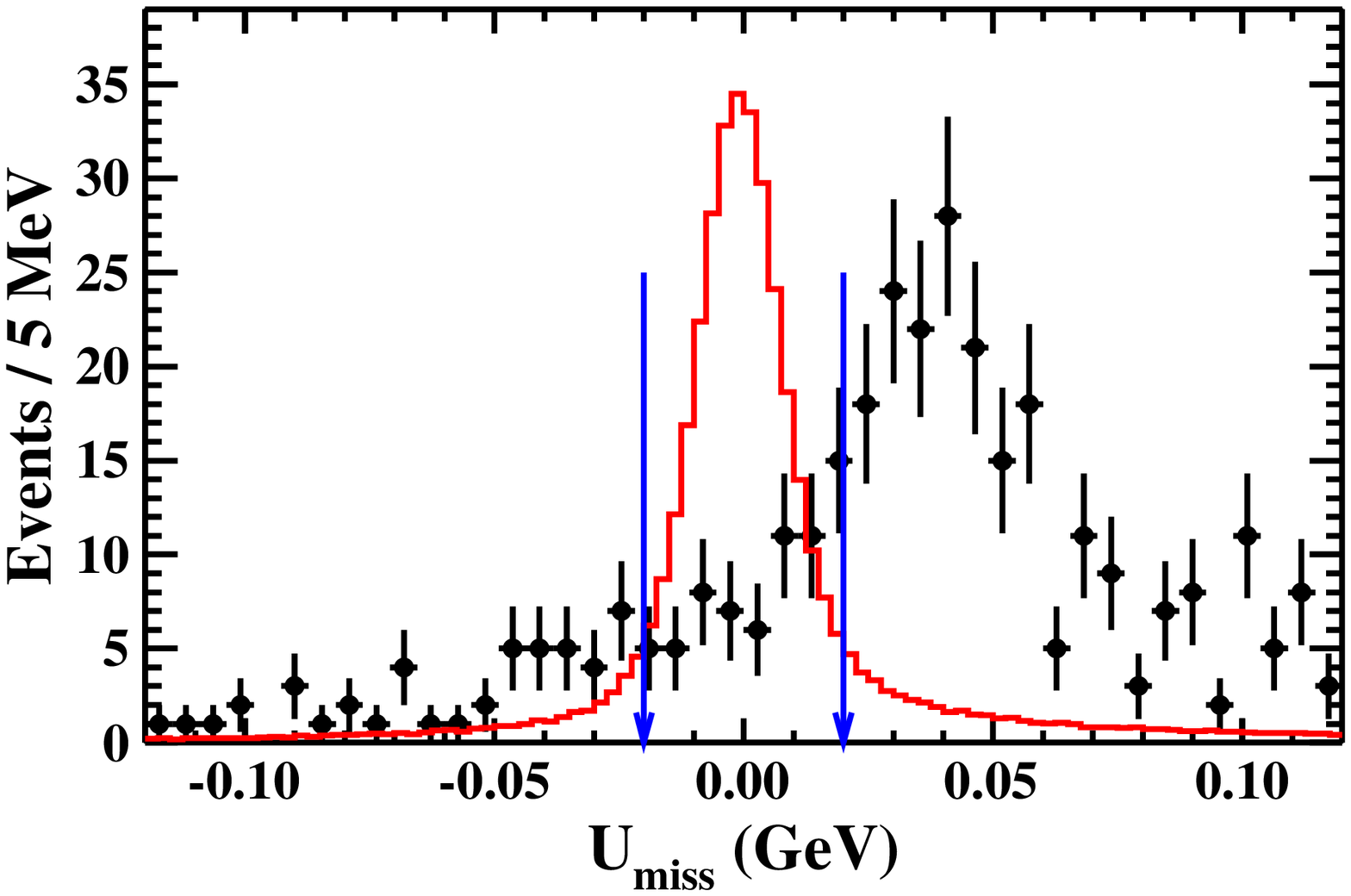}
    \put(-180,120){\bf (b)}}
    \caption{$U_{miss}$ distributions for the decay of (a) $\jpsi \to
    \DsRhom$, and (b) $\jpsi \to \antiDKst$.  The requirements $|
    U_{miss} |< 0.05$\,GeV and $|U_{miss} |< 0.02$\,GeV are
    shown in the figures by vertical arrows.  The dots with error bars are data, while the histograms represent distributions of the arbitrarily normalized exclusive signal MC events.}
  \label{nueu}
\end{figure*}

\begin{figure*}[htbp]
  \centering
  \subfigure{
    \label{recoil_mass:a}
    \includegraphics[scale=0.4]{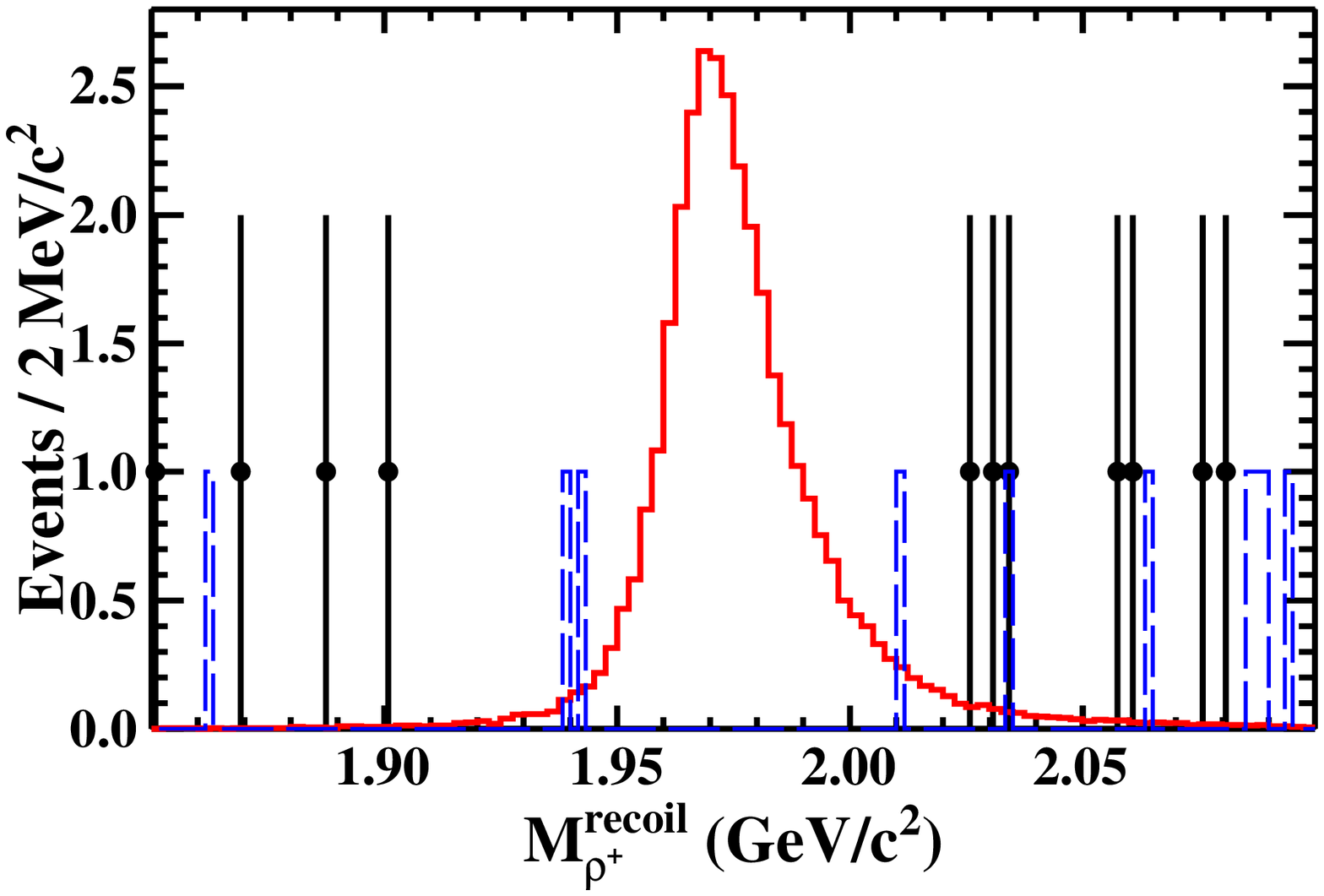}
     \put(-40,125){\bf (a)}}
  \subfigure{
    \label{recoil_mass:b}
    \includegraphics[scale=0.4]{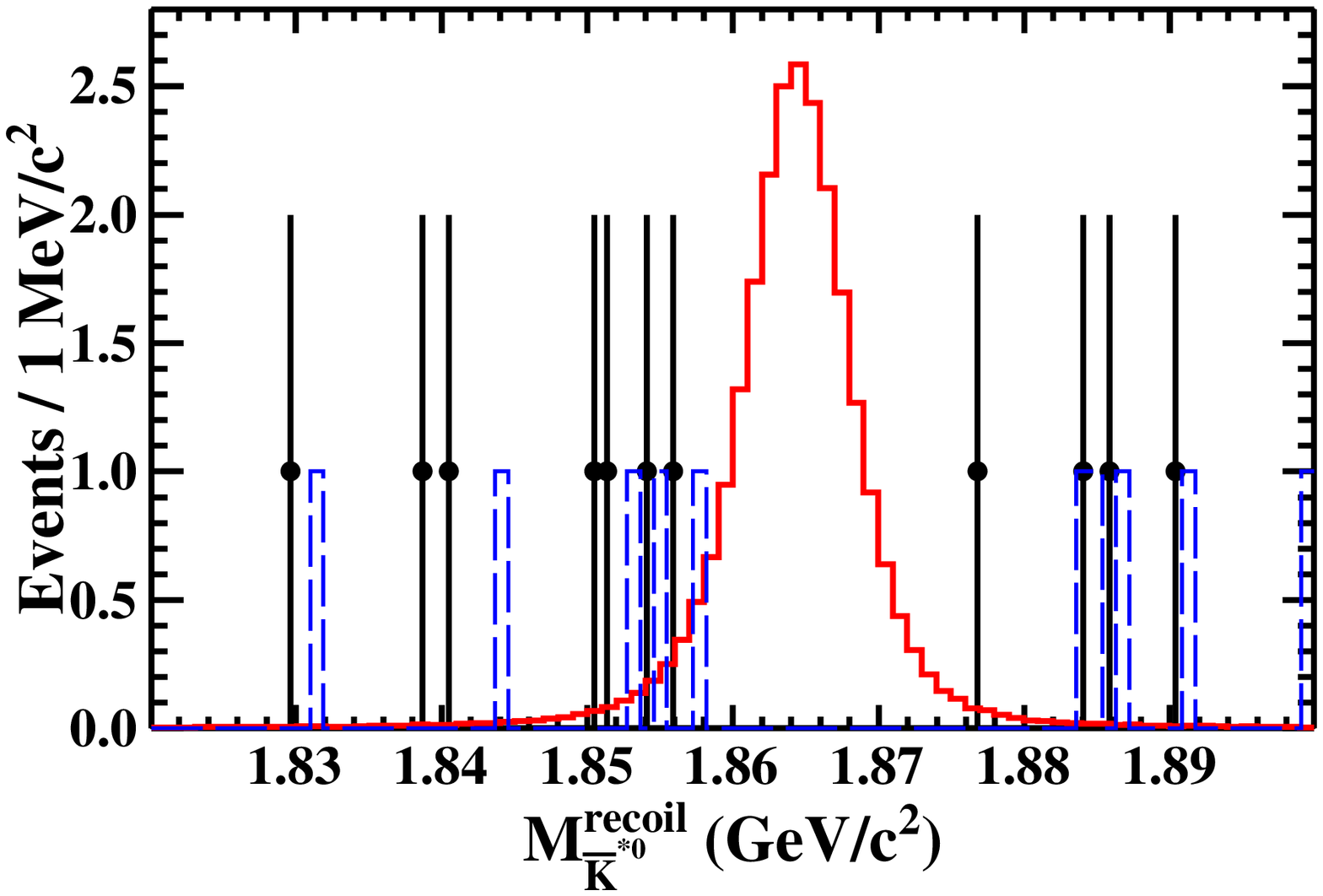}
      \put(-40,125){\bf (b)}}
    \caption{Mass distributions recoiling against (a) $\rho^+$ from
    $\jpsi \to \DsRhom$ and (b) $\overline{K}{}^{*0}$ from $\jpsi \to
    \antiDKst$. Data are shown by dots with error bars. The solid histograms are the unnormalized MC simulated signal events, while the dashed histograms are background distributions from selected inclusive MC events.}

  \label{recoil_mass}
\end{figure*}

The missing four-momentum $(E_{miss}, \vec P_{miss})$, which
represents the four-momentum of the missing neutrino, is determined
from the difference between the net four-momentum of the $\jpsi$
particle and the sum of the four-momenta of all detected particles in
the event. The missing momentum ($P_{miss}$) distribution is shown in
Fig.~\ref{nuep:a}. $P_{miss}$ is required to be larger than 0.1
GeV/$c$ to reduce the backgrounds from $\jpsi$ decays to final states with four charged
particles and no missing particles but with $e/\pi$ misidentification. Figure~\ref{nueu:a} shows the distribution of
$U_{miss} = E_{miss} - cP_{miss}$, and $|U_{miss}|$ is required to
be less than 0.05\,GeV to reduce backgrounds such as
$K^+K^-\pi^+\pi^-$ with multiple $\pi^0$ or $\gamma$ in the final state,
which were not rejected by prior criteria. After all selection
criteria are applied, 11 events survive in the (1.85, 2.10) GeV/$c^2$ mass region in the distribution of mass recoiling against the $\rho^+$, which is shown in Fig.~\ref{recoil_mass:a}. No accumulation of events in the signal region is found.

In the selection of $\jpsi \to \antiDKst \to
K^+K^-\pi^+e^-\overline{\nu}_e$, there are only four charged tracks in
the final state. To suppress backgrounds containing $\pi^0$s,
kinematic fits to the $\pi^0$ mass are also performed if there are at
least two photons in addition to the charged tracks.  If there is
a $\pi^0$ candidate with $\chi^2_{\gamma\gamma} < 20$, the event is
vetoed.  The $K^-\pi^+$ invariant mass distribution is shown in
Fig.~\ref{mass:b}. To select $\overline{K}{}^{*0}$ candidates, the
$K^-\pi^+$ invariant mass is required to satisfy $M_{K^-\pi^+} \in
(0.82,0.98)$\,GeV/$c^2$. The $P_{miss} > 0.1$\,GeV/$c$ and $|
U_{miss} |< 0.02$\,GeV requirements are also used to suppress
the backgrounds with $e/\pi$ misidentification or multiphotons in the
final states, and their distributions are shown in Figs.~\ref{nuep:b}
and \ref{nueu:b}, respectively. After all selection criteria are
applied, 11 events survive in the (1.82, 1.90) GeV/$c^2$ mass region in the distribution of mass recoiling against the $\overline{K}{}^{*0}$, which is shown in Fig.~\ref{recoil_mass:b}. No accumulation of events in the signal region is found.

MC simulations are used to determine mass resolutions, selection
efficiencies and to study possible backgrounds. 600,000 exclusive signal MC
events are generated, and the selection efficiencies are determined to
be $(7.79\pm0.04)\%$ and $(21.83\pm0.06)\%$ for $\jpsi \to \DsRhom$ and $\jpsi \to
\antiDKst$, respectively. 200 million inclusive $\jpsi$ MC events are
used to investigate possible backgrounds from $\jpsi$ decays. For the
decay $\jpsi \to \DsRhom$, 11 MC events pass the final selection
criteria and 8 of them are due to $e/\pi$ misidentification, where a
pion is identified as an electron. For the remaining three events, two
events are $\pi^0 \to \gamma e^+e^-$, where an electron is identified
as a pion, and the other is $\pi^+ \to \mu^+ \nu_{\mu}, \mu^+ \to e^+
\nu_{e}\overline{\nu}_{\mu}$. For the decay $\jpsi \to \antiDKst$, 10 MC events pass the
final selection criteria. Seven events are due to $e/\pi$
misidentification, two events are from $\pi^0 \to \gamma e^+e^-$, and
the other event from $\pi^+ \to \mu^+ \nu_{\mu}, \mu^+ \to e^+
\nu_{e}\overline{\nu}_{\mu}$. From the inclusive MC study, both the number of surviving
background events and their distributions shown as the dashed histogram in Fig.~\ref{recoil_mass}, are consistent with data.

Sideband events are also used to estimate the background. Here, the
backgrounds contributions are estimated using $U_{miss}$ sidebands,
defined as $|U_{miss} |\in (0.05,0.10)$\,GeV and $|U_{miss}
|\in (0.08,0.10)$\,GeV for $\jpsi \to \DsRhom$ and $\jpsi \to
\antiDKst$ respectively. There are 15 and 9 sideband events surviving
in the $D_s^-$ and $\overline{D}{}^0$ mass region. The number of
surviving background events and their distributions from sideband data
are also consistent with data.

\section{Systematic errors}

In this analysis, the systematic errors in the determination of the
branching fraction upper limits mainly come from the following
sources:

\begin{itemize}

\item MDC tracking: The MDC tracking efficiency is studied in clean
channels like $\jpsi \to \rho \pi \to \pi^+ \pi^- \pi^0$, $\jpsi \to
\pppipi$, and $\jpsi \to K_S^0K^+ \pi^-$ samples~\cite{trk_pid_err}. It is found that
the MC simulation agrees with data within 1.0\% for each charged
track. Therefore 4.0\% is taken as the systematic error on the
tracking efficiency for the two channels analyzed with four charged
tracks in the final states.

\item Photon detection: The photon detection efficiency is studied from $\jpsi \to \rho^0\pi^0$ and
  photon conversion via $e^+e^- \to \gamma\gamma$~\cite{pho_err}. The
  difference between the detection efficiencies of data and MC
  simulation is 1.0\% for each photon.

\item Particle ID: The particle ID efficiencies of
electrons, pions, and kaons are studied with samples of radiative
Bhabha events, $\jpsi \to \pppipi$, and $\jpsi \to K_S^0K^+ \pi^-$,
respectively~\cite{trk_pid_err}. The kaon, pion, and electron particle ID efficiencies
for data agree with MC simulation within 1\% for each charged
particle, and 4\% is taken as the systematic error from this source.

\item $\pi^0$ kinematic fit: To estimate the systematic error from the
$\pi^0$ kinematic fit in the analysis of $\jpsi \to \DsRhom$, a clean
$\pi^0$ sample is selected from $J/\psi\rightarrow\rho^+\pi^-(\rho^+
\to \pi^0\pi^+)$ without the kinematic fit. Events with two oppositely
charged tracks identified as pions and two photons
are selected. Further, the $\pi^-$ momentum is required to be in the
range of $P_{\pi^-}\in (1.4,1.5)$\,GeV/$c$, and the
$\pi^+\pi^-\pi^0$invariant mass is required to be in the $J/\psi$ mass
region $|M_{\pi^+\pi^-\pi0}-M_{\jpsi}| < 0.05$\,GeV/$c^2$. In
addition, $E/cP$ is required to be less than 0.8 to remove Bhabha
events.

After the above selection, a same $\pi^0$ kinematic fit as the one in the selection of $\jpsi \to \DsRhom$ is done on the candidates. The same analysis is also performed with MC events. The
efficiency difference between data and MC simulation due to the
$\pi^0$ kinematic fit with $\chi^2<200$ is 0.2\%, which is regarded
as the systematic error.

Applying a similar method, the efficiency difference of the $\pi^0$
kinematic fit used for vetoing events in the decay $\jpsi \to
\antiDKst$ is determined to be $1.0\%$ using a sample of $\jpsi \to
\overline{K}{}^{*0}K^0_S(\overline{K}{}^{*0} \to K^- \pi^+, K^0_S \to
\pi^+\pi^-)$ events.

\item Mass window requirements: The systematic errors of the mass
window requirements are due to the difference in mass resolution between
MC simulation and data and are estimated from some control samples,
which are selected without the mass window requirements. The
uncertainty is obtained by comparing the efficiencies of mass window
requirements between data and MC events. The uncertainties of $\phi$,
$\rho^+$, and $\overline K^{\ast0}$ mass window requirements are
$1.0\%$, $1.0\%$, $0.5\%$ using samples of $\jpsi \to \gamma \phi
\phi(\phi \to K^+K^-) $, $\jpsi \to \rho^+ \pi^-$, and $\jpsi \to
\overline{K}{}^{*0}K^0_S$, respectively.

\item $U_{miss}$ requirement: The systematic error of the
$U_{miss}$ window requirement is due to the mass resolution
difference between MC simulation and data.  Using a similar method as that used for the mass window requirement, the uncertainties of the $U_{miss}$ requirements are $1.0\%$ for $\jpsi \to \DsRhom$ and
$4.0\%$ for $\jpsi \to \antiDKst$, which are different for the two channels since the $U_{miss}$ requirements
are different in these two channels.

\item Intermediate decays: The errors on the intermediate decay
branching fractions of $D_s^- \to \phi e^- \overline{\nu}_e, \phi \to
K^+K^-$, $\rho^+ \to \pi^+\pi^0$,$\pi^0 \to \gamma\gamma$, and $\overline
D^0 \to K^+ e^- \overline{\nu}_e, \overline{K}{}^{*0} \to K^-\pi^+$ are
taken from world average values~\cite{pdg}, and by adding them in
quadrature, $5.7\%$ and $1.1\%$ are the errors for $\jpsi \to \DsRhom$
and $\jpsi \to \antiDKst$, respectively.

\end{itemize}
The systematic error contributions studied above, the error due to
the uncertainty on the number of $J/\psi$ events~\cite{jpsino} and MC statistics are all
summarized in Table~\ref{sys_err}.  The total systematic errors are
obtained by summing them in quadrature, assuming they are independent.

 \begin{table}[htbp]
\centering \caption{Summary of systematic errors ($\%$).}\label{sys_err}
\renewcommand{\arraystretch}{1.2}
\begin{ruledtabular}
\begin{tabular}{ c c c}
Sources&$J/\psi \to \DsRhom$&$\jpsi \to \antiDKst$\\
\hline
MDC tracking &$4.0$&$4.0$\\

Photon detection &$2.0$&$2.0$\\

Particle ID&$4.0$&$4.0$\\

$\pi^0$ kinematic fit&$0.2$&$1.0$\\

$\phi$ mass window&$1.0$&--\\

$\rho^+$ mass window&$1.0$&--\\

$\overline K^{\ast0}$ mass window&--&$0.5$\\

$U_{miss}$ window&$1.0$&$4.0$\\

Intermediate decays&$5.7$&$1.1$\\

MC statistics&$0.5$&$0.3$\\

Number of $J/\psi$ events&$1.2$&$1.2$\\

Total&$8.6$&$7.5$\\

\end{tabular}

\end{ruledtabular}
\end{table}

\begin{table*}[hbtp]
\caption{Numbers used in the calculation of upper limits on the
branching fractions of $\jpsi \to \DsRhom$ and $\jpsi \to
\antiDKst$. $\varepsilon$ is the detection efficiency,
$\mathcal{B}_{inter}$ is the intermediate branching fraction,
$\sigma^{sys}$ is the systematic error, $N_{UL}$ is the upper limit of
the number of observed events at the $90\%$ C.L., $\mathcal{B}$ is the
upper limit at the $90\%$ C.L. on the branching
fraction.}\label{res_sum} \small \renewcommand{\arraystretch}{1.2}
\addtolength{\tabcolsep}{1.0mm}
\begin{ruledtabular}
\begin{tabular}{ c c c c c c c c}
%\hline
Decay mode&Intermediate decay&$\varepsilon$&$\mathcal{B}_{inter}$&$\sigma^{sys}$&$N_{UL}$&$\mathcal{B}$ ($90\%$ C.L.)&\\
\hline
\multirow{2}*{$\jpsi \to \DsRhom$}   &$D_s^- \to \phi e^- \overline{\nu}_e,~\phi \to K^+K^-$,&\multirow{2}*{$7.79\%$}&\multirow{2}*{$1.20\%$}&\multirow{2}*{$8.6\%$}&\multirow{2}*{2.5}&\multirow{2}*{$< 1.3 \times 10^{-5}$}\\

&$\rho^+ \to
\pi^+\pi^0$,~$\pi^0 \to \gamma\gamma$&&& &&&\\

$\jpsi \to \antiDKst$& $\overline{D}{}^0 \to K^+ e^- \overline{\nu}_e,~\overline{K}{}^{*0} \to K^-\pi^+$&$21.83\%$&$2.37\%$&$7.5\%$&2.7&$ < 2.5 \times 10^{-6}$\\

\end{tabular}
\end{ruledtabular}
\end{table*}

\section{Results}
No excess of $\jpsi \to \DsRhom$ or $\jpsi \to \antiDKst$ events above
background is observed. The upper limits on the branching fractions of
these decay modes are calculated using

\begin{equation}
\mathcal{B} < \frac{N_{UL}}{N_{\jpsi}\varepsilon \mathcal{B}_{inter}(1-\sigma^{sys})},
\end{equation}
where $N_{UL}$ is the upper limit of the number of observed events at
the $90\%$ confidence level ($90\%$ C.L.), $N_{\jpsi}$ is the number
of $\jpsi$ events, $\varepsilon$ is the detection efficiency,
$\mathcal{B}_{inter}$ is the intermediate branching fraction, and
$\sigma^{sys}$ is the systematic error.

The upper limits for the observed number of events at the $90\%$
C.L. are 2.5 for $\jpsi \to \DsRhom$ and 2.7 for $\jpsi \to \antiDKst$
using a series of unbinned extended maximum likelihood fits.  In the
fit, the recoil mass distributions of data, shown in
Fig.~\ref{recoil_mass}, are fitted with a probability
density function (p.d.f.) signal shape determined from MC simulations, and the
background is represented by a second-order Chebychev
polynomial. The likelihood distribution, determined by varying the number of signal events from zero to a large number, is taken as the p.d.f. $N_{UL}$ is the number of events corresponding to $90\%$ of the integral of the p.d.f. The fit-related uncertainties are estimated by using different fit ranges and different orders of the background polynomial, and $N_{UL}$ is taken as maximum value among the variations.  All numbers used in the calculations of the upper limits on the branching fractions are shown in Table~\ref{res_sum}.

In summary, a search for the weak decays of $\jpsi \to \DsRhom$ and
$\jpsi \to \antiDKst$ has been performed using a sample of $(225.3 \pm
2.8) \times 10^6 \jpsi$ events collected at the BESIII detector.  No
evident signal is observed and upper limits at the 90\% C.L. are set on
the branching fractions, $\mathcal{B}(\jpsi \to \DsRhom) < 1.3 \times
10^{-5}$ and $\mathcal{B}(\jpsi \to \antiDKst) < 2.5 \times 10^{-6}$,
for the first time. These upper limits exclude new physics predictions which allow flavor-changing processes to occur with branching fractions around $10^{-5}$ but are still consistent with the predictions of the SM.\\

\section{Acknowledgement}
The BESIII collaboration thanks the staff of BEPCII and the computing center for their strong support. This work is supported in part by the Ministry of Science and Technology of China under Contract No. 2009CB825200; Joint Funds of the National Natural Science Foundation of China under Contracts Nos. 11079008, 11179007, 11179014, U1332201; National Natural Science Foundation of China (NSFC) under Contracts Nos. 10625524, 10821063, 10825524, 10835001, 10935007, 11005122, 11125525, 11235011, 11275210; the Chinese Academy of Sciences (CAS) Large-Scale Scientific Facility Program; CAS under Contracts Nos. KJCX2-YW-N29, KJCX2-YW-N45; 100 Talents Program of CAS; German Research Foundation DFG under Contract No. Collaborative Research Center CRC-1044; Istituto Nazionale di Fisica Nucleare, Italy; Ministry of Development of Turkey under Contract No. DPT2006K-120470; U. S. Department of Energy under Contracts Nos. DE-FG02-04ER41291, DE-FG02-05ER41374, DE-FG02-94ER40823, DESC0010118; U.S. National Science Foundation; University of Groningen (RuG) and the Helmholtzzentrum fuer Schwerionenforschung GmbH (GSI), Darmstadt; WCU Program of National Research Foundation of Korea under Contract No. R32-2008-000-10155-0.

\clearpage


\begin{thebibliography}{99}
\bibitem{cc}
The charge conjugate states are implicitly included throughout this paper.
\bibitem{plot}
Throughout this paper, the plot shows the distribution with requirements applied to all other variables except the one shown on it.
\end{thebibliography}
\end{document}